\documentclass[11pt]{article}

\PassOptionsToPackage{dvipsnames}{xcolor}
\usepackage[preprint]{acl}

\usepackage{times}
\usepackage{latexsym}
\usepackage{booktabs}
\usepackage{amsmath}
\usepackage{graphicx}
\usepackage[T1]{fontenc}

\usepackage[utf8]{inputenc}

\usepackage{microtype}
\usepackage{xspace}
\usepackage{xcolor}
\usepackage{seqsplit}

\newcommand{\prism}{\textsc{Prism}\xspace}

\newcommand{\ResearchHub}{\textsc{ResearchHub}\xspace}

\usepackage{inconsolata}

\usepackage{graphicx}
\usepackage{pifont}
\title{\ResearchHub: A Personalized Literature Discovery System for Biomedical Preprints}

\author{
  Viswonathan Manoranjan\textsuperscript{\ding{169}},
  Taeyoon Kim\textsuperscript{\ding{169}},
  Nicholas Do\textsuperscript{\ding{169}},
  Sang Woo Kim\textsuperscript{\ding{169}},
  Dawei Li\textsuperscript{\ding{171}}, \\
  \textbf{Zhen Tan}\textsuperscript{\ding{168}},
  \textbf{Tianlong Chen}\textsuperscript{\ding{169}},
  \textbf{Yunmei Liu}\textsuperscript{\ding{167}} \\
  \textsuperscript{\ding{169}}University of North Carolina at Chapel Hill,
  \textsuperscript{\ding{171}}Arizona State University, \\
  \textsuperscript{\ding{168}}Stevens Institute of Technology,
  \textsuperscript{\ding{167}}University of Louisville
}

\begin{document}
\maketitle
\begin{abstract}
We present \ResearchHub\footnote{Demo: \url{https://ai-4-research.vercel.app/}. Video: \url{https://youtu.be/gjGwHwdYnEk}.}, an interactive system for personalized literature discovery over biomedical preprints. Given a researcher profile or a free-form query, \ResearchHub retrieves relevant papers from a domain-specific MedRxiv corpus and helps users move beyond near-duplicate recommendations through controllable novelty, diversity, and curriculum ordering. At the core of the system is \prism, a bi-encoder retriever adapted to biomedical preprints using GPT-4o-generated natural language queries and a hybrid Multiple Negatives Ranking plus MarginMSE objective with cross-encoder soft labels. On 44,937 MedRxiv papers from 2023--2025, \prism outperforms strong off-the-shelf dense retrievers, improving Recall@10 over SPECTER2-Search by 12.4 percentage points. \ResearchHub further supports frontier retrieval based on a user's reading history, DPP-based diversity re-ranking, Local Intrinsic Dimensionality ordering from mainstream to specialized papers, and lightweight attribution cards for result explanation. The full pipeline runs with low latency, making the system suitable for interactive paper exploration and personalized research monitoring.
\end{abstract}

\section{Introduction}

The volume of scientific papers is growing faster than researchers can read, filter, and organize. This challenge is especially clear in fast-moving biomedical areas, where preprints often appear before formal peer review and can quickly influence new research directions. As a result, literature discovery is shifting from keyword search toward interactive and personalized systems that can model user interests, surface relevant papers, and support lightweight exploration within daily research workflows \citep{takeshita2024gengo,flicke2025scholar}.
\begin{figure*}
    \centering
    \includegraphics[width=1\linewidth]{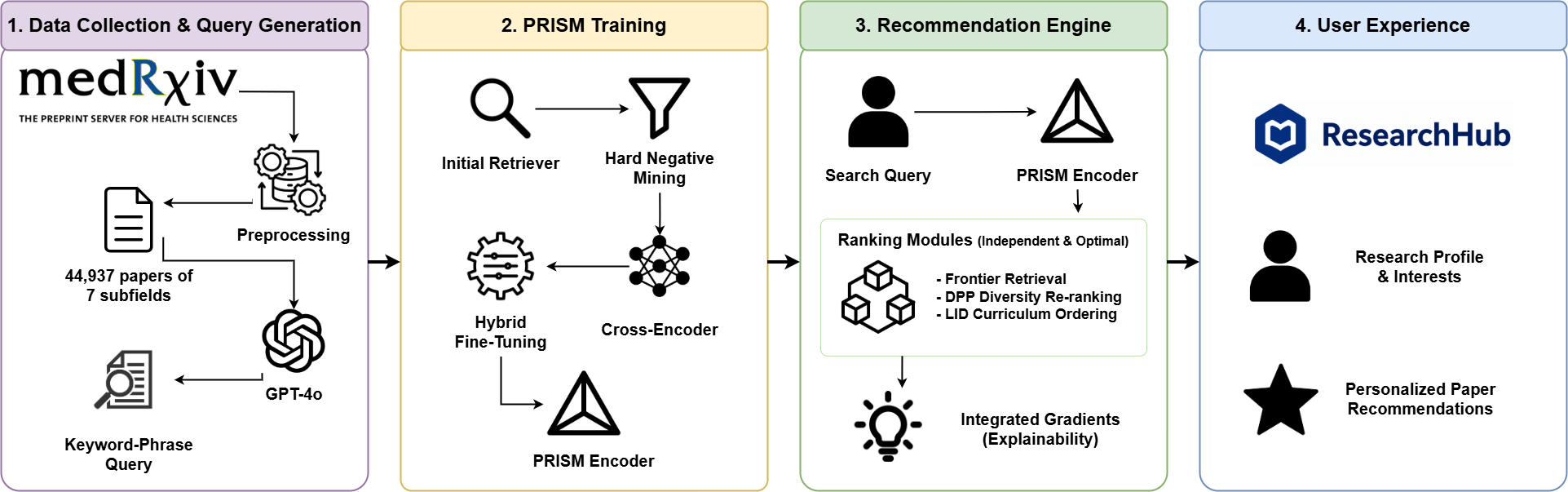}
    \caption{Overview pipeline of \ResearchHub.}
    \label{fig:placeholder}
\end{figure*}
However, current retrieval-based discovery still faces a gap between relevance and exploration. Dense retrievers make large-scale semantic search efficient, and scientific-domain encoders or domain adaptation methods can further improve paper matching quality \citep{karpukhin2020dense,cohan2020specter,wang2022gpl}. Yet a highly relevant ranked list may still be narrow, redundant, or overly close to what the user already knows. For researchers, useful recommendations should not only match their current interests, but also expose adjacent topics, diverse subfields, and frontier papers beyond their existing reading history.

In this paper, we present \textbf{\ResearchHub}, an interactive system for personalized and exploratory biomedical preprint discovery. The system is motivated by a tension in literature search: researchers need recommendations that are close to their interests, but overly close results often become redundant and do little to support discovery. \ResearchHub addresses this tension as a single pipeline. First, \textbf{\prism} (\textit{Preprint Retrieval with Interpretable Semantic Matching}) provides a strong relevance foundation by adapting a bi-encoder to biomedical preprints with GPT-4o-generated natural language queries and a hybrid loss objective. Then, frontier retrieval, Determinantal Point Process (DPP) re-ranking, and Local Intrinsic Dimensionality (LID)-based curriculum ordering reshape the relevant candidate set for exploration.

Frontier scoring moves results away from the user's already-known papers, DPP reduces near-duplicate recommendations by increasing subtopic coverage, and LID ordering presents the final list from mainstream to more specialized work. Lightweight attribution cards help users inspect why papers were retrieved. Together, these components turn dense retrieval from a nearest-neighbor search tool into an interactive discovery workflow that supports both accurate matching and guided exploration \citep{ziegler2005improving,kulesza2012determinantal,amsaleg2018extreme,sundararajan2017axiomatic}.

We evaluate \ResearchHub on the full MedRxiv corpus of 44,937 preprints from 2023 to 2025. \prism substantially improves retrieval quality over strong off-the-shelf baselines, including SPECTER2-Search, with a 12.4 percentage point gain in Recall@10 (the fraction of queries for which the correct paper appears among the top 10 retrieved results). The exploration modules provide controllable trade-offs: frontier retrieval increases novelty, DPP re-ranking improves intra-list diversity and subtopic coverage with limited similarity loss, and LID ordering changes presentation order without changing the retrieved candidate set. System latency remains low enough for interactive use. Overall, \ResearchHub shows that a domain-adapted dense retriever, combined with simple user-controllable exploration tools, can support both precise recommendation and broader literature discovery in biomedical preprints.

\section{Methodology}
\label{sec:method}

\subsection{Corpus and Preprocessing}
\label{sec:corpus}

We build \ResearchHub on MedRxiv preprints posted between January 2023 and November 2025. After deduplication and basic quality filtering, the corpus contains 44,937 papers. For each paper, we retain the title and abstract as the retrieval text. Author names and citation counts are excluded from the retriever input to reduce popularity bias. MedRxiv subject labels are used only for analysis and evaluation, not as retrieval features.


The frontend is built with Next.js and communicates with the backend via RESTful APIs. Authentication is handled through a JWT-based context. The backend is implemented in Flask with PostgreSQL for storage. The recommendation engine uses SentenceTransformers \citep{reimers2019sentence}  to embed papers, indexes them with FAISS \citep{johnson2019faiss} for similarity search, and passes retrieved results to an explainability module that generates per-recommendation reasoning.




\subsection{\prism Retrieval Model}

At the core of \ResearchHub is PRISM, a domain-adapted bi-encoder retriever for biomedical preprints. We start from a general dense retrieval encoder and fine-tune it on synthetic query--paper pairs generated from the MedRxiv corpus. For each abstract, GPT-4o \citep{hurst2024gpt} is prompted to generate concise natural language queries that reflect how a researcher might search for the paper's contribution, rather than simply copying phrases from the abstract. Table~\ref{tab:nlq_examples} shows representative examples of these generated queries alongside their source paper titles.

\prism is trained with a hybrid objective. The Multiple Negatives Ranking (MNR) loss \citep{henderson2017efficient} provides contrastive pressure by pulling each query close to its positive paper and pushing it away from in-batch and mined negatives. MarginMSE \citep{hofstatter2020improving}complements this by distilling graded relevance margins from a cross-encoder teacher, allowing the bi-encoder to preserve relative ranking information among hard candidates. In practice, the two losses play different roles: MNR keeps the embedding space well separated, while MarginMSE improves fine-grained ordering within the candidate set.

\begin{table*}[htbp]
\centering
\footnotesize
\setlength{\tabcolsep}{3pt}
\caption{Recommendation quality on 44,937 NLQ queries.
AvgSimX@$k$ = mean cosine similarity of the top-$k$ retrieved papers
(excluding the ground-truth paper) to the ground-truth paper,
minus the mean random-pair similarity in the embedding space,
thereby normalizing for embedding anisotropy.
Best result in \textbf{bold}; second best \underline{underlined}.}
\label{tab:retrieval}

\begin{tabular}{lccccc}
\toprule
\textbf{Model} &
\textbf{R@10} &
\textbf{R@100} &
\textbf{NDCG@10} &
\textbf{AvgSimX@10} &
\textbf{AvgSimX@50} \\
\midrule

\multicolumn{6}{l}{\textit{Off-the-shelf baselines}} \\

BERT-MS-MARCO &
0.672 &
0.864 &
0.509 &
0.040 &
0.033 \\

DistilBERT-MS-MARCO &
0.668 &
0.855 &
0.512 &
\underline{0.261} &
\underline{0.221} \\

SPECTER2-Proximity &
0.729 &
0.909 &
0.556 &
0.082 &
0.073 \\

SPECTER2-Search &
\underline{0.768} &
\underline{0.927} &
\underline{0.599} &
0.085 &
0.076 \\

\midrule

\multicolumn{6}{l}{\textit{Fine-tuned (this work)}} \\

\textbf{\prism} &
\textbf{0.892} &
\textbf{0.978} &
\textbf{0.739} &
\textbf{0.373} &
\textbf{0.304} \\

\bottomrule
\end{tabular}
\end{table*}





\subsection{Frontier Retrieval}

Standard relevance ranking often returns papers that are close to what the user already knows. To support exploratory discovery, \ResearchHub includes a frontier retrieval mode based on the user's bookmarked papers. Given a candidate paper $p$ and a reading history $H$, we define novelty as the cosine distance from the closest paper in the history:
\[
\mathrm{novelty}(p) = 1 - \max_{k \in H} \cos(e_p, e_k).
\]
The final score interpolates between query relevance and history-based novelty:
\[
s_{\alpha}(p) = (1-\alpha)\cdot \mathrm{sim}(e_q,e_p) + \alpha \cdot \mathrm{novelty}(p),
\]
where $\alpha \in [0,1]$ is controlled by the user. Low values favor conventional relevance, while higher values push the ranking toward papers farther from the user's known set. This gives users a simple way to move from familiar literature toward adjacent or frontier work.




\subsection{Diversity Re-ranking}

A relevant recommendation list can still be redundant if many returned papers discuss nearly the same topic. \ResearchHub therefore includes a DPP-based re-ranking module  \citep{kulesza2012determinantal} over the top candidates returned by FAISS, using the greedy MAP inference algorithm of \citet{chen-etal-2018-fast}. The DPP kernel combines a relevance weight for each candidate with pairwise embedding similarity, encouraging the selected list to contain papers that are both relevant to the query and different from one another. A user-facing parameter $\lambda$ controls this trade-off: higher values preserve relevance more strongly, while lower values increase coverage across subtopics. We use $\lambda=0.5$ as the default setting based on the diversity--similarity trade-off in Section~\ref{sec:results}.





\subsection{Curriculum Ordering with Local Intrinsic Dimensionality}

For users entering a new area, the order of papers can matter. \ResearchHub therefore provides an optional curriculum ordering based on Local Intrinsic Dimensionality (LID) \citep{houle2017local,amsaleg2018extreme}. We pre-compute LID for each paper using its local neighborhood in the embedding space. Papers in dense regions tend to represent mainstream or well-covered topics, while papers in sparse regions often correspond to more specialized or cross-topic work. When the curriculum mode is enabled, the retrieved top-10 papers are sorted by ascending LID, guiding users from central papers toward more frontier papers without changing the underlying candidate set.





\subsection{Explanation Cards}

\ResearchHub presents each recommendation as an explanation card rather than a bare title list. In addition to metadata such as title, authors, year, and abstract snippet, the card highlights query terms that contribute to the dense retrieval representation. We compute these indicative highlights with Integrated Gradients over the \prism query encoder. The highlights are intended to help users understand why a result was retrieved, but we do not treat them as causal explanations of model behavior.

\section{Experiments and Results}
\label{sec:results}
\subsection{Experimental Setup}

We evaluate retrieval on the full corpus of 44,937 MedRxiv papers. Following the GPL-style evaluation protocol, each paper is paired with a held-out natural language query generated from its abstract, and the model is evaluated by retrieving from the full index. We report Recall@10, Recall@100, NDCG@10, MRR, and normalized AvgSimX, which measures the semantic quality of the retrieved neighborhood after subtracting each model's background random-pair similarity.

We compare \prism with four off-the-shelf dense retrieval baselines: BERT-MS-MARCO, DistilBERT-MS-MARCO, SPECTER2-Proximity, and SPECTER2-Search. The first two are general-domain retrieval encoders trained on MS MARCO \citep{bajaj2018msmarcohumangenerated}, while the SPECTER2 variants \citep{singh2023scirepeval} are scientific paper encoders trained with citation-based supervision, building on the original SPECTER model \citep{cohan-etal-2020-specter}. SPECTER2-Search serves as the strongest domain-relevant baseline for ad hoc paper retrieval.



\subsection{Retrieval Benchmarks}

We evaluate using each paper's abstract as a natural language query against the full 44,937-paper FAISS index. Table~\ref{tab:retrieval} shows that \prism achieves the best performance on all metrics. Compared with SPECTER2-Search, the
strongest off-the-shelf baseline by recall, \prism gains +12.4 pp R@10
and +14.0 pp NDCG@10. On neighbourhood quality, \prism leads by a
larger margin: normalised AvgSimX@10 of 0.373 versus 0.085 for
SPECTER2-Search, meaning \prism's top-10 results are semantically
closer to the ground-truth paper by 4.4$\times$ above background.
 
\paragraph{Loss ablation.}
We ablate four configurations: \prism (full system), \prism{} w/o MarginMSE (MNR only), \prism{} w/o MNR (MarginMSE only), and \prism{} w/o NLQ (T5 queries with hybrid loss).
Table~\ref{tab:loss_ablation} shows that each component contributes distinctly.
MNR prevents embedding-space collapse and preserves neighborhood quality.
MarginMSE provides the graded supervision needed to rank the target paper highly.
GPT-4o NLQ queries improve both retrieval and neighborhood quality over T5 keyword fragments.
Removing any single component degrades performance, confirming that the components are complementary.
\begin{table}[t]
\centering
\small
\setlength{\tabcolsep}{4pt}
\caption{Loss \& query ablation (44,937 NLQ queries).
N@10 = NDCG@10; AX = normalised AvgSimX (minus background $\bar{s}$,
see Appendix~\ref{par:avgsimx_norm}).
$\dagger$~collapse ($\bar{s}{=}0.88$).
\textbf{Bold} best; \underline{underlined} 2nd.}
\label{tab:loss_ablation}
\begin{tabular}{lcccc}
\toprule
\textbf{Model} & \textbf{R@10} & \textbf{N@10} & \textbf{AX@10} & \textbf{AX@50} \\
\midrule
\textbf{\prism}  & \textbf{0.892} & \textbf{0.739} & \underline{0.373} & \underline{0.304} \\
w/o MNR           & \underline{0.841} & \underline{0.684} & 0.044$^\dagger$ & 0.037$^\dagger$ \\
w/o MarginMSE     & 0.820 & 0.646 & \textbf{0.451} & \textbf{0.373} \\
w/o NLQ (T5)      & 0.804 & 0.640 & 0.168 & 0.142 \\
\bottomrule
\end{tabular}
\end{table}

\begin{figure}[htbp]
\centering
\includegraphics[width=\columnwidth]{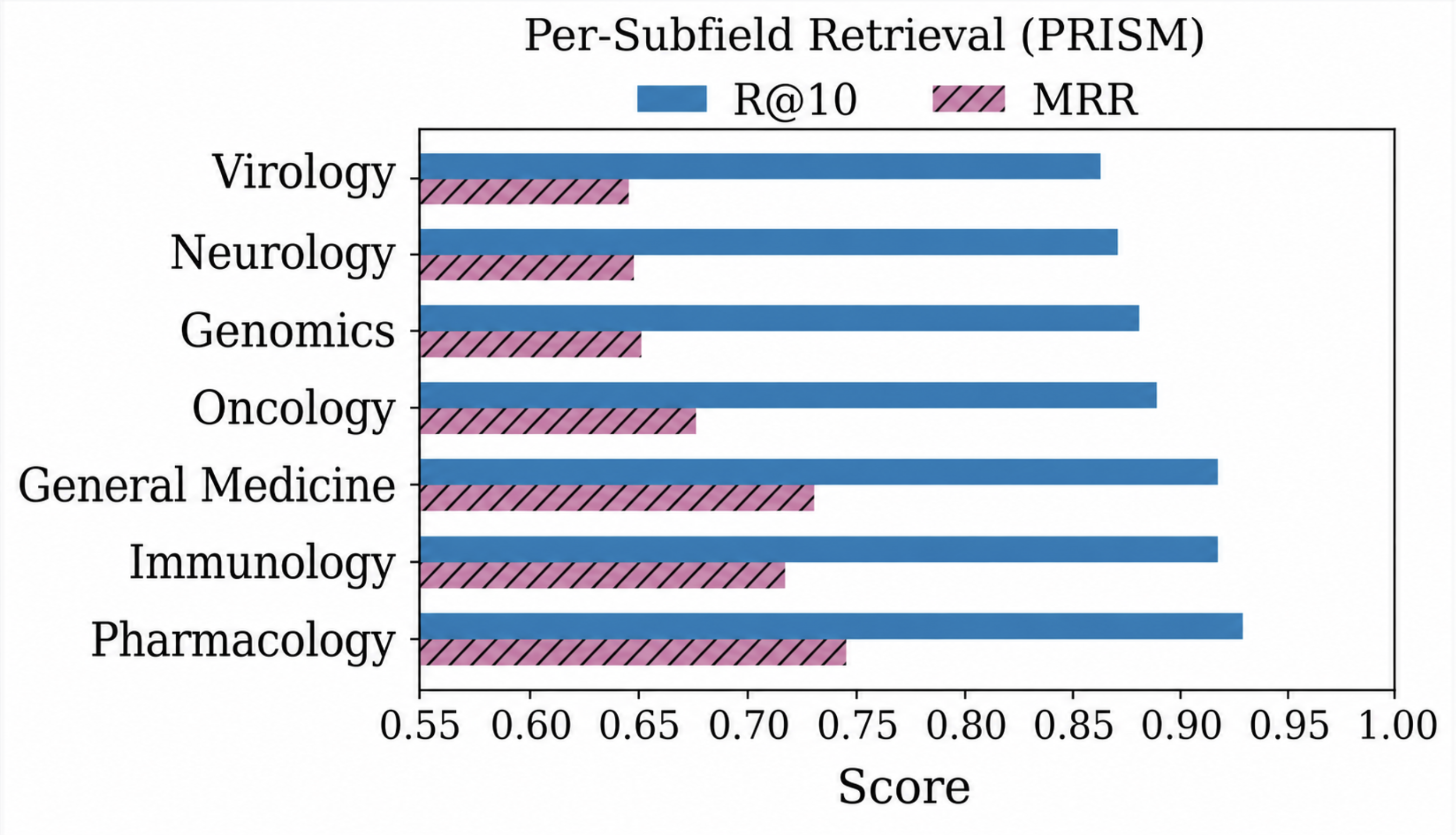}
\caption{Per-subfield Recall@10 and MRR for \prism. Bars sorted by R@10 descending.}
\label{fig:subfield}
\end{figure}

\begin{figure}[t]
\centering
\includegraphics[width=\columnwidth]{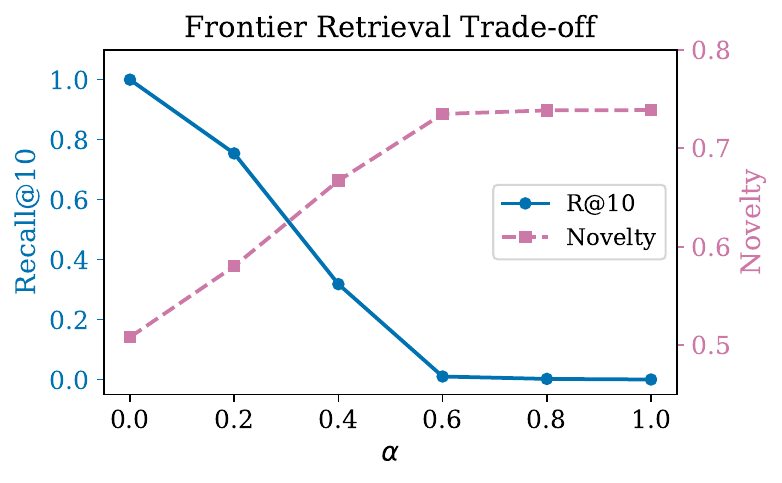}
\caption{Frontier retrieval trade-off. Recall@10 (left axis) decreases as $\alpha$
increases, while novelty (right axis, mean cosine distance from known papers) rises.
Recency ($\approx$0.71 throughout) confirms exploration does not regress to older literature.}
\label{fig:frontier}
\end{figure}

\begin{figure}[t]
\centering
\includegraphics[width=\columnwidth]{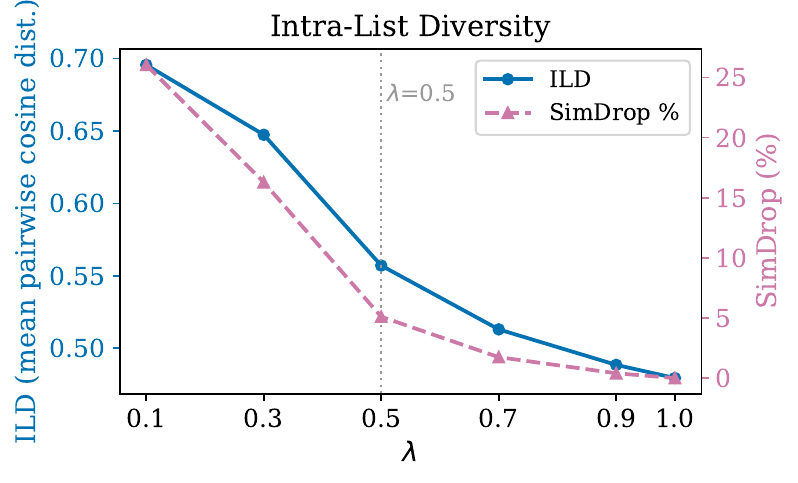}
\caption{DPP sweep: intra-list diversity (ILD) and similarity drop (SimDrop) vs.\ $\lambda$
over 2,000 queries. Dotted line marks the recommended default $\lambda{=}0.5$.}
\label{fig:dpp_ild}
\end{figure}

\begin{figure}[htbp]
\centering
\includegraphics[width=\columnwidth]{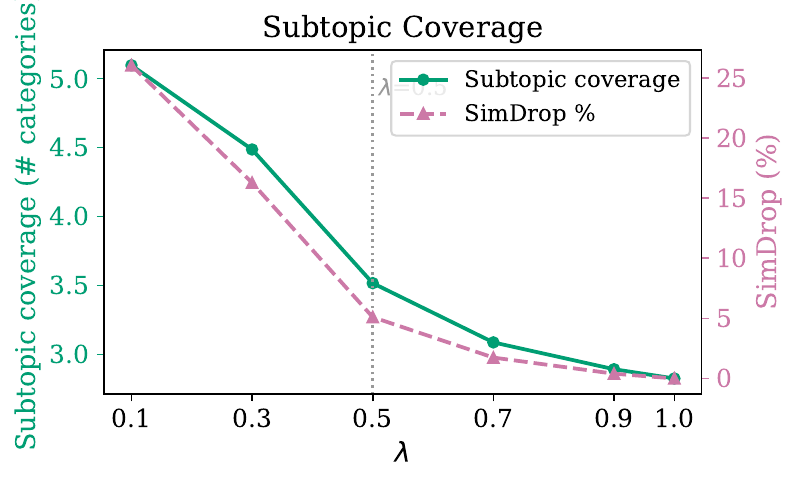}
\caption{DPP sweep: subtopic coverage vs.\ $\lambda$ over 2,000 queries. Recall@10 is
flat at 0.876 across all $\lambda$ (not shown).}
\label{fig:dpp_topics}
\end{figure}

\paragraph{Subfield analysis.}
Figure~\ref{fig:subfield} shows per-topic Recall@10 across the seven
largest MedRxiv topics. Performance is consistently strong, with
Pharmacology (0.928) and Immunology (0.916) highest and Virology
(0.855) lowest, suggesting that virology abstracts span a broader
range of pathogens and methodologies, making retrieval harder.

\subsection{Frontier Retrieval}

We simulate reading histories by sampling 500 corpus papers, using
their 5 nearest neighbors as ``known'' papers and the 6th neighbor
as held-out ground truth. Figure~\ref{fig:frontier} shows the
Recall--Novelty trade-off as $\alpha$ increases.

At $\alpha{=}0.2$, Recall@10 is 0.754 while novelty rises
14\% ($0.508 \to 0.580$). At $\alpha{=}0.4$, novelty increases 31\%
with Recall@10 at 0.318. Recency is stable throughout ($\approx$0.71),
confirming that exploration does not regress toward older literature.
We recommend $\alpha \in [0.2,\,0.4]$ as the interactive default for
researchers seeking cross-subfield discovery.

\subsection{DPP Diversity}

Figures~\ref{fig:dpp_ild} and~\ref{fig:dpp_topics} show the DPP sweep over 2,000 queries.
Recall@10 is flat at 0.876 across all $\lambda$, confirming that
DPP re-ranking does not remove relevant papers from the candidate
pool. At $\lambda{=}0.5$, ILD increases 16\% ($0.479 \to 0.557$)
and subtopic coverage grows from 2.83 to 3.52 topics per list
at only a 5.1\% mean-similarity drop, which falls within the 10\%
validity threshold. We set $\lambda{=}0.5$ as the system default.

\begin{figure*}[t]
    \centering
    \includegraphics[width=0.8\textwidth]{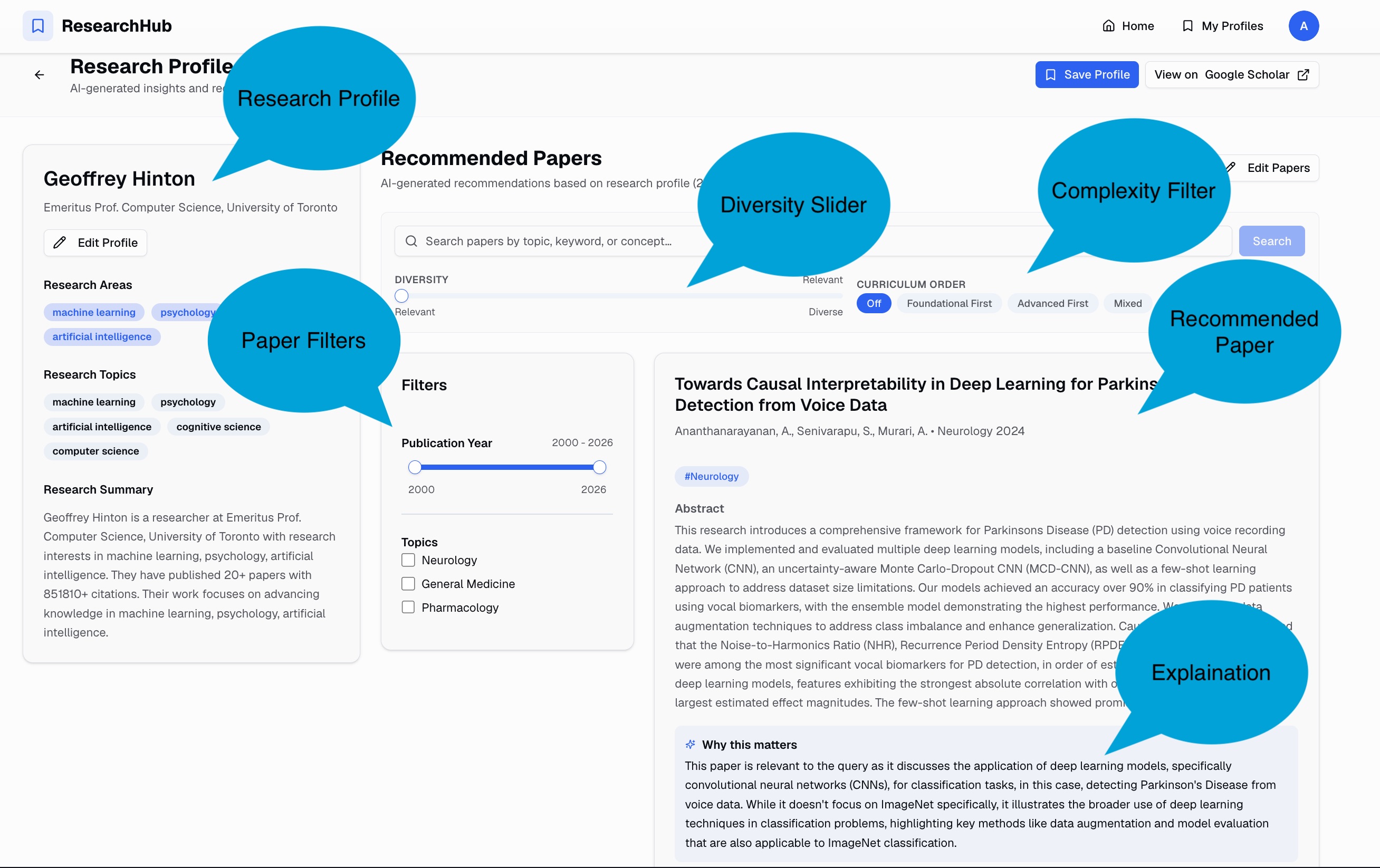} 
    \caption{A screenshot exhibits how the application presents one research profile and recommended research papers}
    \label{fig:ui_screenshot}
\end{figure*}

\subsection{LID Curriculum Ordering}

\paragraph{Candidate-set integrity.}
LID ordering reranks the already-retrieved top-10 results and does
not alter the underlying FAISS candidate pool. We confirm this over 5,000 queries and find that Recall@200 (the
candidate pool before any re-ranking) is 1.00 in all cases
(\texttt{parity\_pass=true}), verifying that the LID sort cannot
introduce false negatives. Because Recall@$k$
measures set membership and not position within the set, all
reported Recall and MRR figures are unaffected by LID reordering
by construction.

\paragraph{Topical centrality interpretation.}
The LID ordering reflects geometric structure in embedding space,
placing papers surrounded by many similar abstracts first and
geometrically isolated papers last. This corresponds to a
\emph{topical centrality} ordering, progressing from mainstream
well-represented subfields toward niche frontier work. Over the full
44,937-paper index ($k{=}20$), LID ranges from 0.15 to 68.8 with
mean 8.14 and a right-skewed distribution where the sparse high-LID
tail consists predominantly of papers in emerging or cross-disciplinary
topics. Very short or boilerplate abstracts account for the extreme
tail. The UI caps the displayed LID indicator at the 95th-percentile
value to suppress these outliers.

\subsection{System Latency}

End-to-end query latency (GPU encode + FAISS search) is
\textbf{10.8\,ms} at p50 and 11.5\,ms at p95 for single queries.
Throughput reaches 5.5\,ms per query at batch size 10. DPP re-ranking adds $<$2\,ms.
Frontier novelty scoring (max-cosine scan over $|\mathcal{H}|{\leq}50
\times 200$ pairs, NumPy vectorised) adds $<$0.3\,ms. LID-based
curriculum sort over 10 pre-computed values adds $<$0.1\,ms.
The 131\,MB FAISS index fits in CPU memory with no GPU required
at inference, enabling lightweight deployment.

\subsection{Interpretability Analysis}

To characterise where the model's retrieval signal comes from,
we run Integrated Gradients (IG) attribution on 500 queries. We mask
the top-5 attributed tokens and compare Recall@10 against an equal-size
random mask. Attribution masking drops Recall@10 from 0.898 to 0.542
($\Delta{=}{-}0.356$), while random masking yields $\Delta{=}{-}0.486$.
Attribution masking causes \emph{less} degradation than random masking, indicating that retrieval signal is broadly distributed across query tokens rather than concentrated in a few keywords. This is expected for dense encoders that embed holistic meaning. Token-level attributions are shown in the UI as indicative highlights, not causal explanations.

\subsection{Application User Interface}

A user submits a researcher profile URL and the backend retrieves their publication data to build a structured research profile. As shown in Figure~\ref{fig:ui_screenshot}, the frontend displays a dashboard with the researcher’s name, institution, research areas, and an AI-generated summary.

Recommendations appear as interactive cards showing the paper title, authors, venue, year, and per-token attribution highlights from the IG module indicating which query terms drove retrieval. Users can tune novelty exploration via the frontier $\alpha$ slider, control result diversity via the DPP $\lambda$ slider, and toggle LID-based curriculum ordering to progress from mainstream to specialized literature. They can also filter results by year or topic, open the original paper, or use the Contact Authors feature to generate a personalized outreach email. The full pipeline from query encoding through retrieval, re-ranking, and attribution runs end-to-end with no manual steps required.

\section{Conclusion}

We presented \ResearchHub, an interactive system for personalized and exploratory biomedical preprint discovery. The system combines a domain-adapted dense retriever, \prism, with user-controllable modules for novelty, diversity, curriculum ordering, and lightweight explanation. Experiments on 44,937 MedRxiv preprints show that \prism improves retrieval quality over strong off-the-shelf baselines, while the exploration modules help users move beyond near-duplicate recommendations with low latency overhead. These results suggest that effective literature discovery systems should support both accurate matching and guided exploration. In future work, we plan to extend \ResearchHub to broader scientific domains, incorporate implicit user feedback, and evaluate the system with researchers in real-world literature discovery workflows.

\section*{Limitations}
\prism is currently scoped to MedRxiv preprints from 2023--2025 and does not cover peer-reviewed publications or other scientific domains. Extending to a broader corpus requires re-running the GPL fine-tuning pipeline, which depends on GPT-4o for query generation and is not trivially reproducible without API access.

On the personalization side, frontier novelty is computed relative to bookmarked papers and users with sparse reading histories receive weaker signals. The system does not incorporate implicit feedback such as click-through, so recommendations do not improve with use.

Two of our evaluation metrics have interpretive caveats. AvgSimX@$k$ is confounded by local corpus density, so papers in sparse topical neighborhoods score lower regardless of model quality. LID can conflate topic breadth with embedding sparsity from very short abstracts.

\bibliography{custom}
\appendix

\section*{Appendix}
\label{sec:appendix}

\section{Methodology}
\label{app:training}

\subsection{Representative \prism examples}

\begin{table}[htbp]
\centering
\small
\setlength{\tabcolsep}{4pt}
\begin{tabular}{p{0.57\linewidth} p{0.36\linewidth}}
\toprule
\textbf{Paper title} & \textbf{\prism{} query} \\
\midrule
Excess Mortality and Years of Potential Life Lost Among Black People in the US from 1999 to 2020
  & impact of heart disease on Black mortality rates \\
\addlinespace
Design and Retrospective Cohort Validation of an Algorithm for Surgical Site Infection
  & multivariable algorithm for surgical site infection validation \\
\addlinespace
Perceived Seizure Risk in Epilepsy --- Chronic Electronic Surveys with and without Mood
  & perceived seizure risk and mood in epilepsy patients \\
\addlinespace
A Noninvasive Clinical Method to Measure In Vivo Mechanical Strains of the Lamina Cribrosa
  & glaucoma medication effect on lamina cribrosa strain \\
\bottomrule
\end{tabular}
\caption{Representative \prism{} query examples. GPT-4o generates
concise keyword-phrase queries reflecting the paper's core topic,
mirroring how researchers naturally search for related work.}
\label{tab:nlq_examples}
\end{table}

\subsection{GPL Fine-Tuning with T5 Queries (GPL-v4 Baseline)}

The T5-based baseline uses the BeIR T5-large query generator
(\texttt{\seqsplit{query-gen-msmarco-t5-large-v1}}) \citep{nogueira2019doc2query,thakur-etal-2021-beir} to generate five synthetic
queries per abstract (\(224{,}685\) total; four per document for training, one
held out), yielding \(179{,}748\) training queries and a \(44{,}937\)-query
test set. Queries are filtered for length (15--60 tokens) and language, with
312 non-English outputs discarded and regenerated. We fine-tune
\texttt{\seqsplit{msmarco-bert-base-dot-v5}} with MultipleNegativesRankingLoss
over cross-encoder-mined hard negatives
(\texttt{\seqsplit{ms-marco-MiniLM-L-6-v2}} \citep{wang2020minilm}, top-50 candidates, top-5 pairs
retained per query; see Appendix~\ref{app:hard_negative}), for 2 epochs at
batch size 32, \(\text{lr}=2\times10^{-5}\), warmup ratio 0.1
(\(\approx 42{,}128\) steps, single GPU).

\subsection{Cross-Encoder Scoring and Hard Negative Mining}
\label{app:hard_negative}

For each query, we retrieve the top-100 candidates via FAISS and score every
query--document pair with a cross-encoder
(\texttt{\seqsplit{ms-marco-MiniLM-L-6-v2}}), $r(q,d) \in [0,1]$. To avoid
false negatives, hard negatives are sampled from candidates ranked 5--50
(excluding the top-5), and the top-5 cross-encoder-scored pairs per query are
retained for training.

\subsection{Natural Language Query (NLQ) Generation and Training}

Beyond T5-generated queries, we explore domain adaptation using natural language queries
generated by the OpenAI API (GPT-4o), which produce more diverse and realistic
search phrasings compared to the template-style T5 output.

\paragraph{NLQ query generation.}
For each of the \(44{,}937\) medRxiv papers, we generate five distinct natural language queries
using the OpenAI API, yielding \(224{,}685\) queries in total.
Queries are stored per document and cover a range of phrasings: short keyword-style queries
(e.g., \textit{``excess mortality rates among Black people in the US 1999 to 2020 study''}),
citation-style queries, and longer descriptive questions.

\paragraph{Training objectives.}
Section~\ref{sec:method} describes the respective roles of MNR and MarginMSE in the hybrid objective. Two implementation details are worth adding here: the cross-encoder is used only for hard-negative mining and pair selection and does not appear in the MNR loss itself, and MarginMSE alone (without MNR's contrastive pressure) lets the embedding space collapse, since it constrains only relative similarity margins rather than absolute distances between embeddings -- this is the mechanism behind the $\dagger$-marked collapse in Table~\ref{tab:loss_ablation}.

\begin{table*}[htbp]
\centering
\small
\setlength{\tabcolsep}{4pt}
\caption{Background cosine similarity ($\bar{s}$, mean random-pair cosine sim
on 2,000 sampled documents) and normalised AvgSimX@$k$ for all models.}
\label{tab:background_sim}
\begin{tabular}{lccccc}
\toprule
\textbf{Model} & $\bar{s}$ & \textbf{Raw@10} & \textbf{Norm@10} & \textbf{Raw@50} & \textbf{Norm@50} \\
\midrule
BERT-MS-MARCO        & 0.871 & 0.911 & 0.040 & 0.904 & 0.033 \\
DistilBERT-MS-MARCO  & 0.283 & 0.545 & 0.261 & 0.504 & 0.221 \\
SPECTER2-Proximity   & 0.839 & 0.921 & 0.082 & 0.912 & 0.073 \\
SPECTER2-Search      & 0.847 & 0.932 & 0.085 & 0.923 & 0.076 \\
\prism{} w/o NLQ     & 0.595 & 0.761 & 0.167 & 0.733 & 0.138 \\
\textbf{\prism}      & \textbf{0.021} & \textbf{0.393} & \textbf{0.373} & \textbf{0.325} & \textbf{0.304} \\
\bottomrule
\end{tabular}
\end{table*}

\paragraph{NLQ GPL training configuration.}
We fine-tune a GPL model using the NLQ queries with the following setup:
\begin{itemize}
  \item \textbf{Base encoder}: sentence-transformers/msmarco-bert-base-dot-v5.
  \item \textbf{Query source}: OpenAI-generated natural language queries
        (\(5\) per document; \(4\) used for training, \(1\) held out for evaluation).
  \item \textbf{Training queries}: \(179{,}748\) (4 per document used for training,
        1 held out per document for evaluation), yielding \(898{,}740\) training pairs
        (\(5\) top cross-encoder pairs per query).
  \item \textbf{Cross-encoder}: \textbf{cross-encoder/ms-marco-MiniLM-L-6-v2}
        scoring top-50 candidates per query.
  \item \textbf{Training}: MNR + MarginMSE hybrid loss, \(2\) epochs,
        batch size \(= 32\), learning rate \(= 2\cdot 10^{-5}\),
        warmup ratio \(= 0.1\), \(\approx 14{,}042\) total training steps
        (using 4-GPU DDP, effective batch \(= 128\)).
\end{itemize}
The held-out test set contains \(44{,}937\) queries (one per document).

\subsection{Similarity Diagnostics}

Beyond standard ranking metrics, we compute similarity-based diagnostics to
understand how semantically close the retrieved documents are to the ground truth.

\textbf{AvgSimX@k} (as defined in Table~\ref{tab:retrieval}) isolates the
semantic quality of the surrounding recommendations independent of whether
the exact target paper was retrieved. In the main results table, we include normalised AvgSimX@10 and AvgSimX@50, which summarize
the semantic tightness of the neighbourhood around the ground-truth document after
subtracting each model's background similarity (see Appendix~\ref{par:avgsimx_norm}).
The raw AvgSim@$k$ values (without background subtraction) are shown in
Table~\ref{tab:background_sim} for reference.

\paragraph{Normalised AvgSimX@k.}
\label{par:avgsimx_norm}
Raw AvgSimX values are not directly comparable across models. Different training
objectives produce embedding spaces with different degrees of
\emph{anisotropy}, the tendency of all embeddings to cluster in a narrow region
of the unit sphere. In a highly anisotropic space, any two documents will have
a high cosine similarity simply by default, so a model can score a high raw
AvgSimX@$k$ not because it retrieves semantically relevant neighbours but
because everything is close to everything.

To account for this, we sample 2,000 random corpus documents per model and
compute the mean pairwise cosine similarity, which we call the
\emph{background similarity} $\bar{s}$. Normalised AvgSimX@$k$ is then:
\[
\widetilde{\mathrm{AvgSimX}}@k = \mathrm{AvgSimX}@k - \bar{s}
\]
This tells us how much more similar the retrieved neighbourhood is to the
ground-truth paper compared to a randomly chosen document, capturing genuine
semantic signal above the noise floor.

Table~\ref{tab:background_sim} shows background similarities and normalised
AvgSimX@10 and @50 for all models. SPECTER2 and BERT-MS-MARCO have background
similarities between 0.84 and 0.87, leaving very little signal above the noise
floor, with normalised AvgSimX@10 ranging from 0.040 to 0.085. \prism is trained
with a hybrid MNR + MarginMSE loss; MNR actively pushes unrelated documents apart,
giving a background of only 0.021. This low background is what allows
\prism's raw AvgSimX@10 of 0.393 to translate into a normalised score well
above the off-the-shelf baselines, as reported in Section~\ref{sec:results}.

\end{document}